%% file: main.tex
\documentclass[twocolumn]{aastex631}

\usepackage{xspace}

\newcommand{\spender}{\textsc{spender}\xspace}

\newcommand{\bx}{\mathbf{x}}
\newcommand{\bs}{\mathbf{s}}
\newcommand{\bw}{\mathbf{w}}
\newcommand{\pgal}{p_{\rm galaxy}}

\begin{document}
\title{Outlier Detection in the DESI Bright Galaxy Survey}

\correspondingauthor{Yan Liang}
\email{yanliang@princeton.edu}

\author[0000-0002-1001-1235]{Yan Liang}
\affiliation{Department of Astrophysical Sciences, Princeton University, Princeton, NJ 08544, USA}

\author[0000-0002-8873-5065]{Peter Melchior}
\affiliation{Department of Astrophysical Sciences, Princeton University, Princeton, NJ 08544, USA}
\affiliation{Center for Statistics \& Machine Learning, Princeton University, Princeton, NJ 08544, USA}

\author[0000-0003-1197-0902]{ChangHoon Hahn}
\affiliation{Department of Astrophysical Sciences, Princeton University, Princeton, NJ 08544, USA}

\author[0000-0001-6662-7306]{Jeff Shen}
\affiliation{Department of Astrophysical Sciences, Princeton University, Princeton, NJ 08544, USA}

\author[0000-0003-4700-663X]{Andy Goulding}
\affiliation{Department of Astrophysical Sciences, Princeton University, Princeton, NJ 08544, USA}

\author[0000-0002-4557-6682]{Charlotte Ward}
\affiliation{Department of Astrophysical Sciences, Princeton University, Princeton, NJ 08544, USA}



\begin{abstract}
We present an unsupervised search for outliers in the Bright Galaxy Survey (BGS) dataset from the DESI Early Data Release. This analysis utilizes an autoencoder to compress galaxy spectra into a compact, redshift-invariant latent space, and a normalizing flow to identify low-probability objects. The most prominent outliers show distinctive spectral features such as irregular or double-peaked emission lines, or originate from galaxy mergers, blended sources, and rare quasar types, including one previously unknown Broad Absorption Line system. 
A significant portion of the BGS outliers are stars spectroscopically misclassified as galaxies. 
By building our own star model trained on spectra from the DESI Milky Way Survey, we have determined that the misclassification likely stems from the Principle Component Analysis of stars in the DESI pipeline. To aid follow-up studies, we make the full probability catalog of all BGS objects and our pre-trained models publicly available.
\end{abstract}



\section{Introduction}
\label{sec:intro}
Large spectroscopic surveys, such as the Sloan Digital Sky Survey~\citep[SDSS;][]{york2000sloan}, have gathered spectroscopic data for millions of astronomical objects. The next generation of surveys, such as the Dark Energy Spectroscopic Instrument \citep[DESI;][]{DESI_Collaboration2016-ht}, produce even larger datasets.
While the majority of cataloged objects can be classified using known spectral types,  there also exist``unknown unknowns''---rare phenomena or object types that deviate from established physical models. 
In such cases, unsupervised machine learning methods can identify features in the data without relying on predefined labels or pre-determined templates. 
Such methods are particularly valuable for detecting outliers, which are, by definition, rare and often unexpected.

In this work, we search for outliers in the Early Data Release~\citep[EDR;][]{desi_edr} of the DESI Bright Galaxy Survey~\citep[BGS;][]{bgs_ts}, which we briefly describe in \autoref{sec:data}. We employ the spectrum autoencoder (AE) architecture \spender introduced by \citet{melchior2022}. This technique compresses galaxy spectra into a compact, redshift-invariant latent space, representing the ``type'' of galaxies, and makes full use of the observed spectra across all redshifts. Following the approach of \citet{liangoutliers}, we interpret the sample density in the AE latent space as a probability distribution, and identify outliers as low-probability objects using a normalizing flow (NF) model. We describe this approach in \autoref{sec:method}.

Our previous application of this method to SDSS galaxy spectra uncovered a range of astrophysical and instrumental outliers, including blends of multiple galaxies and/or stars, extremely reddened galaxies, as well as stars that had been misclassified as galaxies \citep{liangoutliers}. With the new DESI dataset, we again expect outliers to correspond to galaxies in unusual physical states.
We also expect them to reveal failures or artifacts of the data processing pipelines and therefore serve as an alternative mechanism for quality control, sensitive to problems that were not even known to be problems.

In \autoref{sec:outliers}, we present and discuss the nature of remarkable outliers in the BGS galaxy sample, including merging galaxies and rare quasars. 
Our visual inspection also reveals a substantial fraction to be misclassified stars.
In \autoref{sec:star-galaxy}, we extend our generative modeling approach to stellar spectra from the DESI Milky Way Survey~\citep[MWS;][]{mws_ts}, and identify the likely reason why some BGS galaxies were misclassified.
We conclude in \autoref{sec:conclusion} with a discussion of the limitations and potential extensions of this study.

\section{Data}
\label{sec:data}
\input{data}

With those cuts, we compile approximately 250,000 BGS and 210,000 MWS spectra.
The observed wavelength range is $\lambda_{\rm obs}=3600 \dots 9824$\AA. 
We split the samples into training, validation, and testing sets, each accounting for 70\%, 15\%, and 15\% of the entire samples, respectively. 
We apply a mask to the top $\sim$100 telluric lines, assigning zero weights to all bins within 5\,\AA\ of the lines centers, which amounts to approximately 25\% of the data vectors. The spectra are then normalized by the median flux over the rest-frame wavelengths $\lambda_{\rm rest}=5300\dots5850$\AA, a region that is relatively quiescent and accessible at all redshifts, thereby avoiding redshift-dependent encoding.

\section{Methods}
\label{sec:method}

\begin{figure*}[t]
    \centering
    \includegraphics[width=\textwidth]{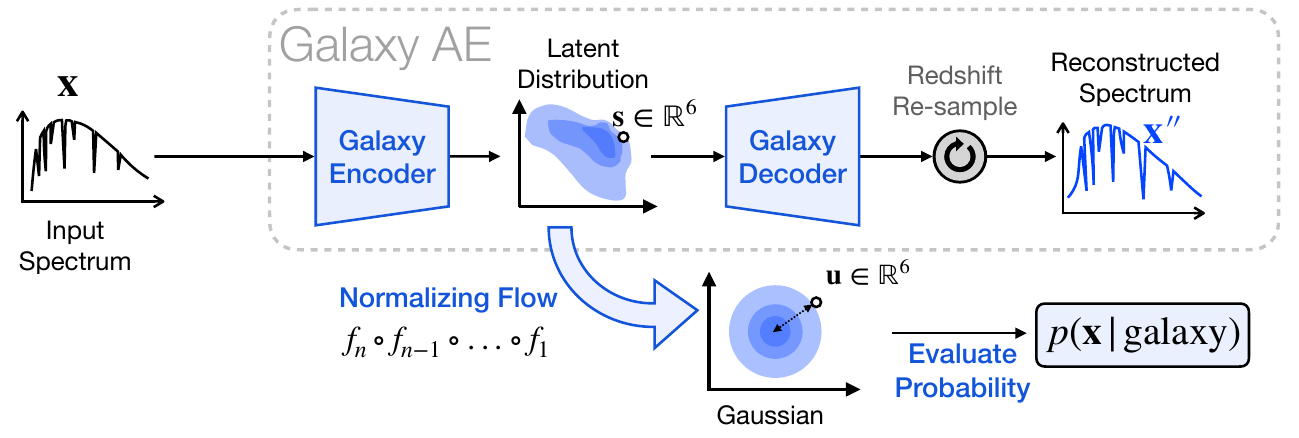}
    \caption{Our approach combines an autoencoder (AE) for spectrum reconstruction and a normalizing flow (NF) for probability estimation. The AE compresses an observed galaxy spectrum into a low-dimensional latent representation of the restframe spectrum, and the NF transforms the latent distribution of the entire BGS galaxy sample into a Gaussian base distribution for efficient probability computation and sampling.}
    \label{fig:diagram}
\end{figure*}

Our general approach follows \citet{liangoutliers} and is summarized in \autoref{fig:diagram}.
In short, we train the spectrum autoencoder architecture \spender \citep{melchior2022} on the observed BGS spectra.
That means, we encode a spectrum $\bx\in\mathbb{R}^M$ into a small number of latent variables $\bs\in\mathbb{R}^S$. For this work, we set $S=6$.
From these latents, the decoder produces a restframe model $\bx'$, which is redshifted and resampled to produce an observed-frame reconstruction $\bx''$ that matches the observation.
In the first training phase, we adjust the network weights so as to minimize the fidelity loss, which quantifies the reconstruction quality over batches of $N$ spectra:
\begin{equation}
L_{\rm fid} = \frac{1}{2NM}  \sum_i^{N}  \bw_i\odot(\bx_i-\bx_i''(\bx_i,z_i))^2,
\end{equation}
where $z_i$ is the known redshift of source $i$, $\mathbf{w}_i$ is the inverse variance vector, and $\odot$ the element-wise multiplication.
For Gaussian-distributed noise, this loss measures the mean log-likelihood of data given the autoencoder model.

In the second phase, we train with the fidelity loss and two extra losses that make the latent space distribution approximately redshift-invariant.
This property yields a more physically meaningful latent distribution and has proven effective for finding outliers in SDSS spectra \citep{liangoutliers}.
The first extra loss term ensures that spectra that are close in data space are also close in latent space; the second one promotes proximity in latent space between the original spectrum and an augmented version, where we modified the redshift and added additional noise to the spectrum.

\begin{figure*}[t]
    \centering
    \includegraphics[width=1.03\textwidth, trim=1em 0 1em 0]{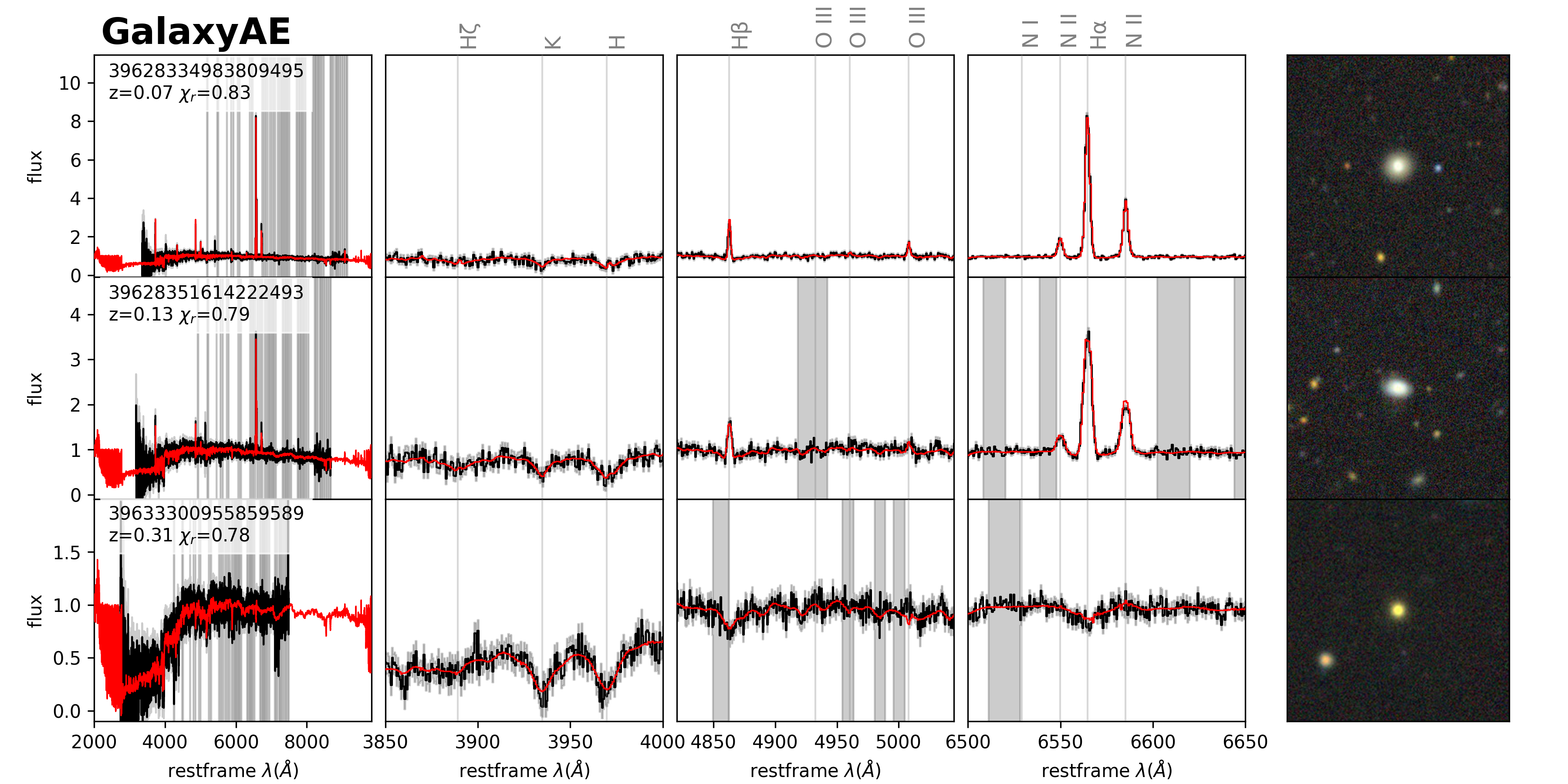}
    \caption{BGS galaxy spectra (black) and their reconstructed models (red) as determined by our spectrum autoencoder. The first column presents the entire flux-normalized restframe spectrum, assuming the {\em Redrock} redshift, with greyed out regions being masked during training. The second, third, and fourth columns provide zoomed-in views of significant emission/absorption lines. The final column shows the image of the source from the DESI Legacy Imaging Surveys.}
    \label{fig:fidelity}
\end{figure*}

Upon convergence of our two-phase training, we achieved fidelity losses of 0.384, 0.391, and 0.386 on the training, validation, and test sets, respectively. The consistency across different datasets indicates that our model is generalizing well on unseen datapoints, with no evidence of significant overfitting. The achieved fidelity corresponds to a reduced chi-square ($\chi^2$) value of 0.77. 
\autoref{fig:fidelity} presents the reconstructed restframe models for three representative galaxy spectra in BGS.
As intended, the model also exhibits redshift invariance. In the latent space of the trained auto-encoder model, original spectra and their artificially redshifted augmentation spectra are on average 21 times closer to each other than the average pairwise distance between different objects. This indicates that our model successfully captures the intrinsic characteristics of galaxies, regardless of their redshift.

As last step, we train a specific type of normalizing flow~\citep[NF;][]{tabak2010density,tabak2013family} model known as a Masked Autoregressive Flow~\citep[MAF;][]{papamakarios2017masked} to describe the density of galaxies in the autoencoder latent space. The combination of an AE and a NF is synergistic: The NF provides a flexible density estimator for the complex latent distributions AEs tend to form, while the AE serves as a non-linear compression scheme to provide a lower-dimensional space for the NF to operate in.
Because of our extended training procedure, the latents are approximately invariant under changes of redshift, rendering them much more informative conditioning variables to describe intrinsic galaxy properties than the observed spectrum data vector.
With the AE-NF combination, we can evaluate the probability $\pgal=p(\bx\mid\mathrm{BGS})$ of any spectrum $\bx$ to be drawn from the BGS sample.
Spectra with low $\pgal$ can thus be interpreted as outliers in the galaxy sample.

\section{BGS Outliers}
\label{sec:outliers}

\begin{figure*}[t]
    \centering
    \includegraphics[width=1.03\textwidth, trim=1em 0 1em 0]{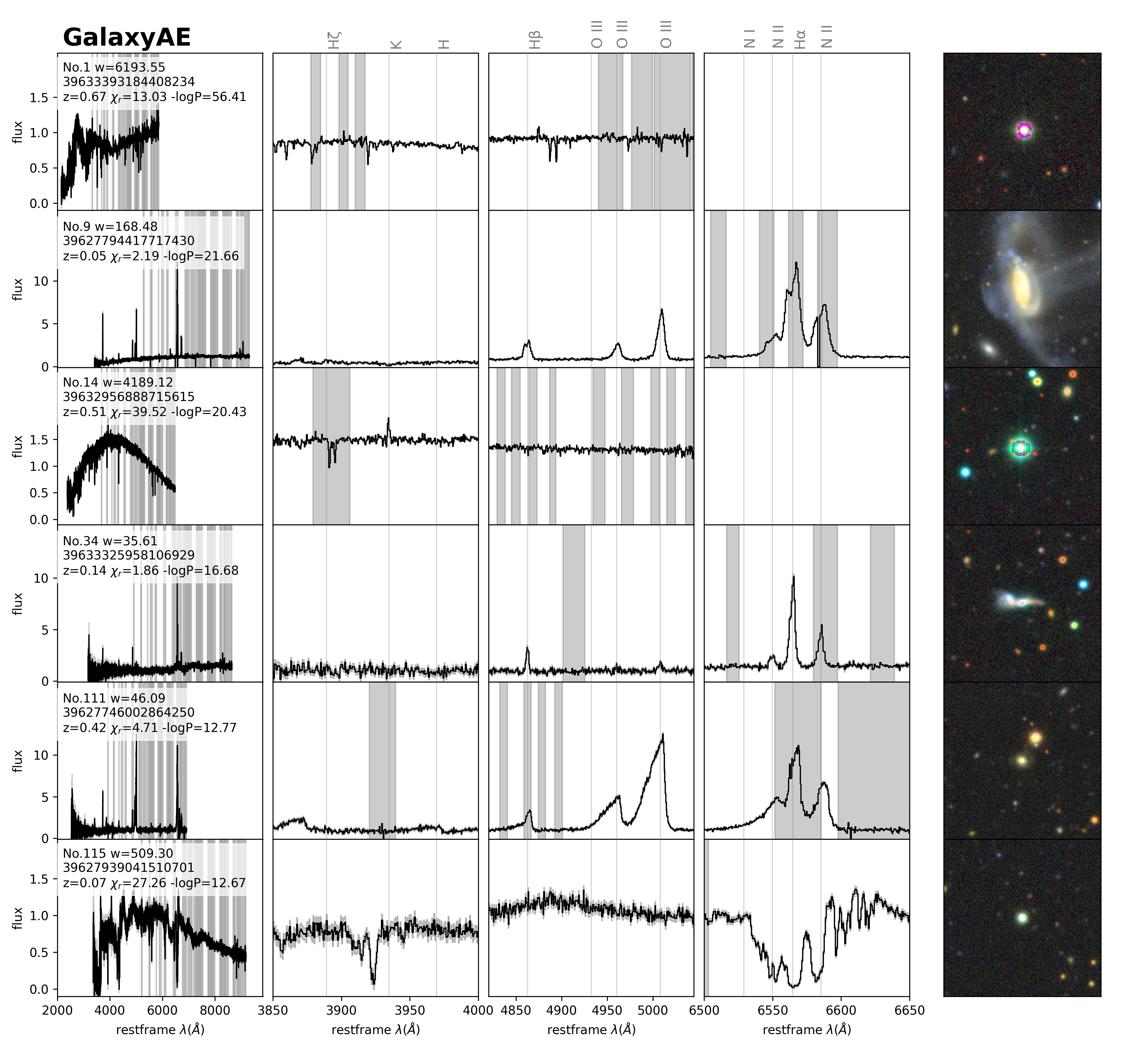}
    \caption{Representative selection of outliers in BGS identified by the galaxy autoencoder and normalizing flow, in decreasing order of the outlier score (details for each object in \autoref{sec:outliers}). The panels are identical to \autoref{fig:fidelity}, but for visual clarity we omitted showing the AE model. The legend in the upper left corner includes the DESI Target ID, {\em Redrock} redshift, reduced chi-square of the AE model, and the negative log probability of the NF model.}
    \label{fig:outliers}
\end{figure*}

We define an outlier score as $-\log\pgal$ and visually investigate the top 200 outliers. 
\autoref{fig:outliers} shows a selection of them.
A large fraction contain irregularly--shaped or double-peaked narrow emission lines (e.g. outlier 9\footnote{DESI 39627794417717430}, shown in the 2nd row of \autoref{fig:outliers}), indicating that the spectra contain overlapping galaxies in cluster environments or galaxy mergers.
Outlier 20\footnote{DESI 39627788356946460} is reported as a galaxy merger in \citet{DePropis2007}, outlier 11\footnote{DESI 39627896888755461} is an ultraluminous infrared galaxy in a merger \citep{Kim2002} with prominent polycyclic aromatic hydrocarbon (PAH) emission \citep{lai2020all}, and outliers 59\footnote{DESI 39628362636858151} and 60\footnote{DESI 39633407830920247} also have morphologies consistent with a merger. Outliers 4\footnote{DESI 39633149726034475}, 43\footnote{DESI 39633402051166367}, 51\footnote{DESI 39627782338122510}, 
91\footnote{DESI 39627794400937953}, 107\footnote{DESI 39633127701745009}, 125\footnote{DESI 39627800390405432}, 133\footnote{DESI 39633154075526488}, 135\footnote{DESI 39633419134568709} and 191\footnote{DESI 39627781759307042} have double-peaked narrow emission lines. Double-peaked narrow emission lines can arise due to complex kinematics in the narrow-line region surrounding a single AGN induced by outflows or disk rotation, or the presence two separate narrow-line regions associated with two merging AGN at small relative velocities \citep{Shen2011}. In the cases of 107 and 125, there is a galaxy merger visible in imaging data, but the others have settled galaxy morphologies and no obvious mergers, making them candidates for close separation dual AGN in the late stages of merger. Outlier 151\footnote{DESI 39627746069975548} has a triple-peaked narrow line due to a cluster environment apparent in imaging data. 

A fraction of the top 200 outliers are reported as dusty, obscured and/or radio-loud AGN in the literature, including outlier 60\footnote{DESI 39633407830920247} \citep{Koziel-Wierzbowsk2020}, 69\footnote{DESI 39627776273159194} \citep{Chen2018}, and 84\footnote{DESI 39628334988006041} \citep{Best2012}.
Outlier 69 has previously been classified as a Wolf-Rayet galaxy due to detectable emission line features from a large population of Wolf-Rayet stars \citep{Chen2018}. Outlier 37\footnote{DESI 39627734065874069; TNS AT2020dig} was reported on the Transient Name Server \citep{AT2020dig} due to the detection of AGN--like optical variability in time-domain imaging from the Zwicky Transient Facility \citep{Bellm2019,Graham2019}. 

Previously unreported, outlier 111\footnote{DESI 39627746002864250} (5th row in \autoref{fig:outliers}) has unusually blue-skewing emission lines indicating the presence of high-velocity outflows, similar to those observed in extremely red quasars \citep[ERQs;][]{Ross2015, Zakamska2016}. Although this object's broad-band color $r-\mathrm{W4} = 18.96-6.744 = 12.22$ is not quite as extreme as required by these earlier works ($r-\mathrm{W4}>14$, where W4 refers to the Vega magnitude in the WISE W4 filter), it is also fainter and towards the low end of the redshift range covered earlier. In any case, line shapes and photometric properties suggest strongly accelerated gas with high levels of dust obscuration or reprocessing in the interstellar medium of the host.

Also previously unknown, outlier 115\footnote{DESI 39627939041510701} (last row in \autoref{fig:outliers}) has complex and broad absorption features.
With the {\em Redrock} redshift estimate of $z=0.0698$, the H$\alpha$ region would be highly unusual.
However, we believe that this source is in fact a Broad Absorption Line \citep[BAL;][]{Hall2002} quasar at $z=1.535$, showing  MgII$\lambda2800$, CIII]$\lambda1909$, and CIV$\lambda1549$ absorption that removes almost all flux directly blueward of the nominal line centers.
If true, it demonstrates the ability to search for these highly unusual quasars in DESI BGS at fainter luminosities than were accessible in SDSS.

We also find a variety of blended objects, similar to SDSS galaxy outliers we reported in \cite{liangoutliers}. Outliers 8\footnote{DESI 39627823501017204}, 12\footnote{DESI 39633145246517940}, 21\footnote{DESI 39632991718212824}, and 53\footnote{DESI 39633416341160166} are each consistent with a superposition of two stars. Outlier 34\footnote{DESI 39633325958106929} (4th row in \autoref{fig:outliers}) shows a chance alignment of an M-type star and a distant galaxy merger at $z=0.13$. Outlier 26\footnote{DESI 39633123197062964} are multiple galaxies, and outlier 15\footnote{DESI 39627782380067362} is a pair of quasars.

However, among the top 200 outliers in the BGS galaxy sample, the largest group consists of 35 stars, including outlier 1\footnote{DESI 39633393184408234} (1st row of \autoref{fig:outliers}).
Like outlier 14\footnote{DESI 39632956888715615} (3rd row of \autoref{fig:outliers}), they often show no indication of blending and are bright enough to reveal the diffraction spikes and saturate the LS images.
We suspect that their brightness led to biases in the imaging measurements, including the estimation of size, which made these object appear extended and thus targeted in BGS.
What surprises us is that {\em Redrock} classified these sources as galaxies with sufficient confidence ({\tt DELTACHI2 > 40}), sometimes even placing them at cosmological redshifts.

\section{Stellar Spectrum Model}
\label{sec:star-galaxy}

\begin{figure*}[t]
    \centering
    \includegraphics[width=\linewidth]{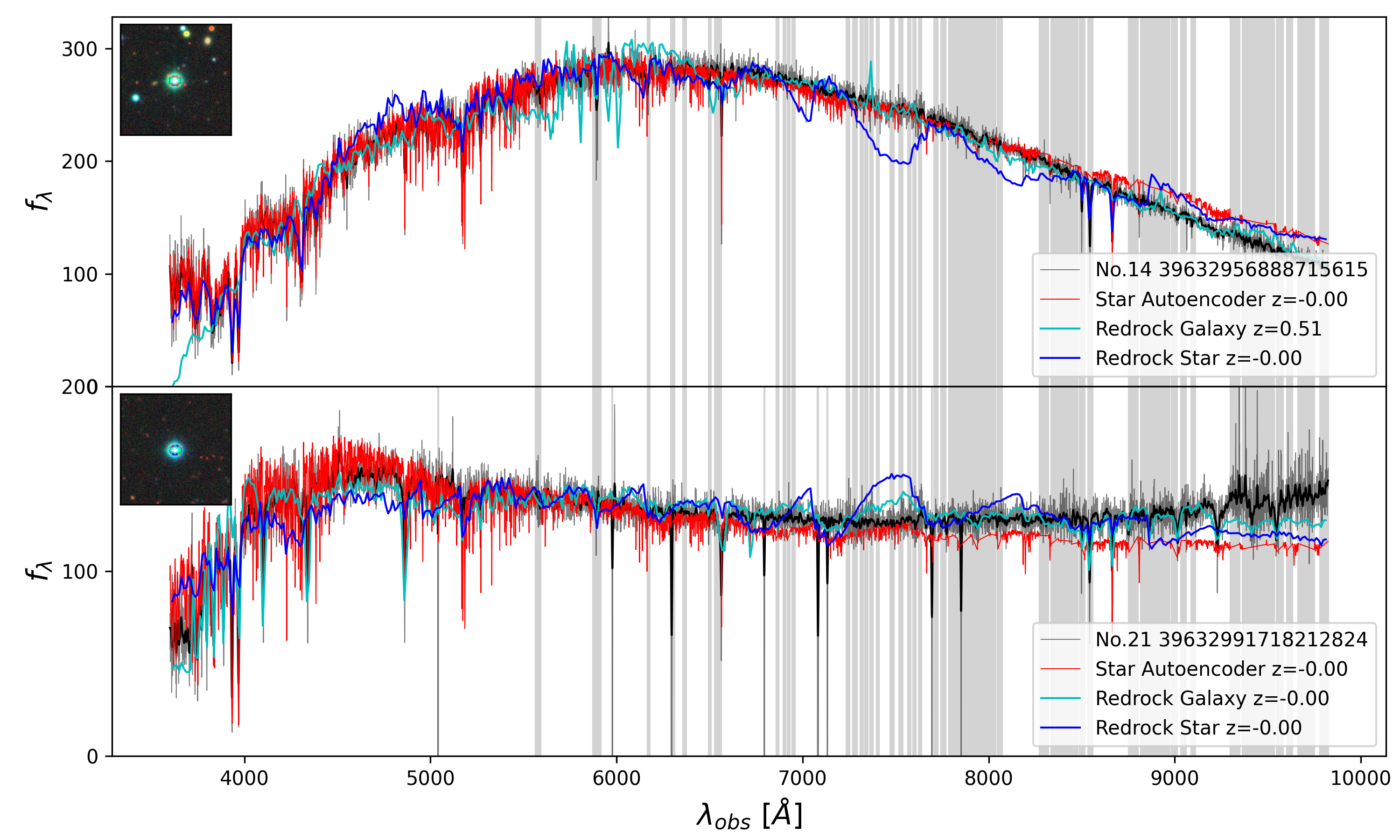}
    \caption{Comparison of observed spectra (black) and model fits ({\em Redrock} galaxy model (cyan), {\em Redrock} star model (blue), and our star autoencoder model (red)) for two example stars misclassified as galaxies by {\em Redrock}. Shaded regions indicate areas masked due to telluric contamination.}
    \label{fig:star-example}
\end{figure*}

We further investigate the stellar contamination in the galaxy sample.
A closer inspection of these stellar outliers shows that {\em Redrock} finds a plausible fit with its galaxy PCA templates, but shows larger residuals for its star templates (see \autoref{fig:star-example}).
These residuals resemble the prominent metal oxide (TiO, VO) absorption features in M dwarfs, but the residuals flip sign in different sources. 
As these features cannot arise in emission, we surmise that either the mean PCA spectrum or the eigenvectors {\em Redrock} uses for its star fits were trained with an overabundance of M dwarfs, so that these metal oxide bands get imprinted on the models of other stars. 

\begin{figure*}[t]
    \centering
    \includegraphics[width=1.0\textwidth, trim=1em 0 1em 0]{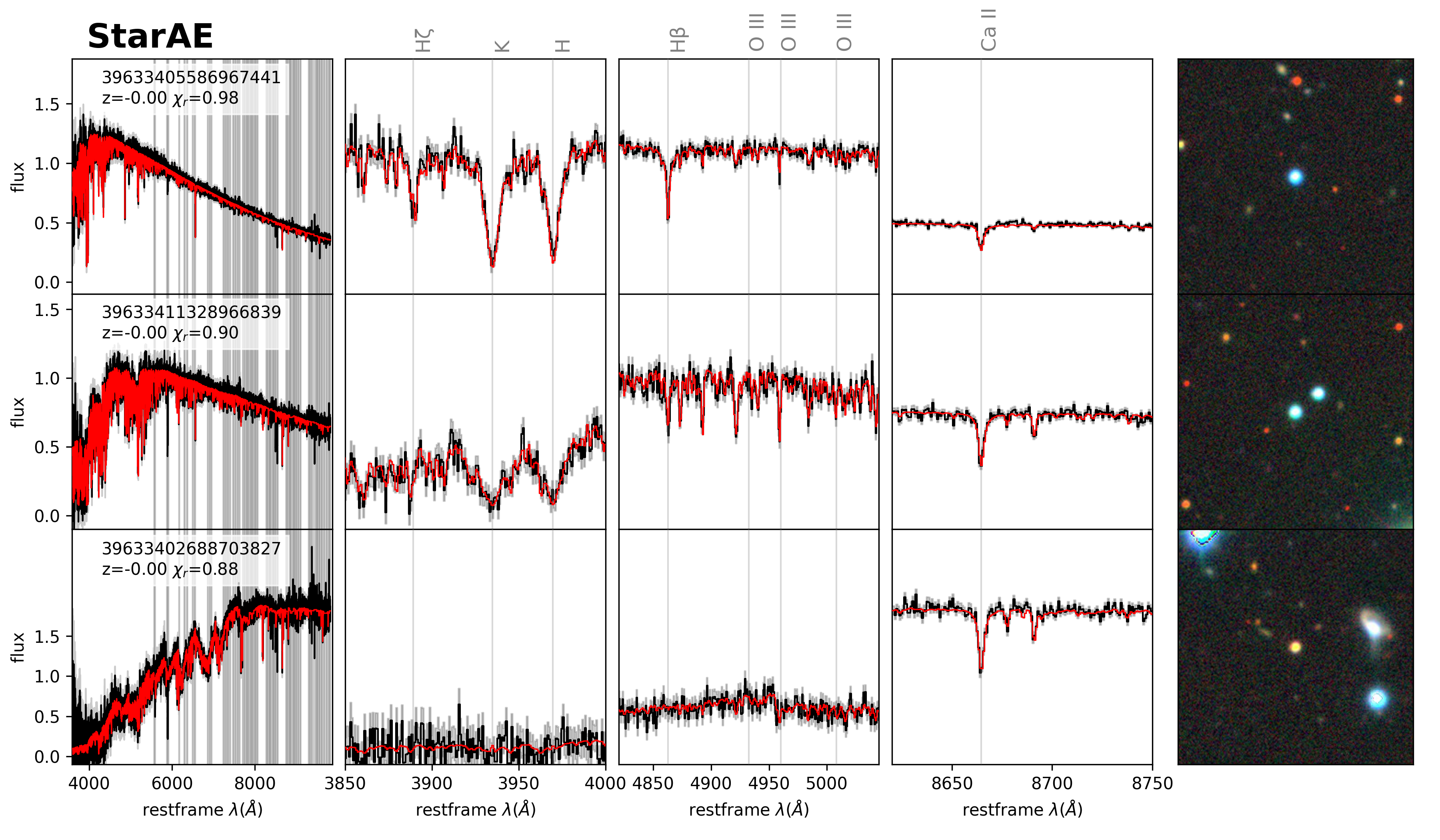}
    \caption{DESI MWS spectra (black) and their reconstructed models (red) from our star autoencoder. Panels as in \autoref{fig:fidelity}.}
    \label{fig:star-fidelity}
\end{figure*}

To test this hypothesis we make our own star model by following the generative modeling approach of \autoref{sec:method} and repeat the autoencoder and NF training with stars in the DESI MWS.
The typical quality of the StarAE reconstructions of a representative selection of MWS stars can be seen in \autoref{fig:star-fidelity}.
We also include the StarAE model in \autoref{fig:star-example}, demonstrating that it is not affected by the spurious features of the PCA model.
We conclude that \spender successfully deals with the diversity in stellar types, variations in radial velocity, and observational imperfections, and produces an accurate model that is not limited to a linear combination of orthogonal templates.

We also repeat the training for a NF on the MWS latent distribution to find additional outliers in the BGS sample, following the suspicion that certain targets do not conform neatly to either star or galaxy class.
For example, outliers 5\footnote{DESI 39633158374688229}, 22\footnote{DESI 39633278931569179}, and 23\footnote{DESI 39632981354087949} are confirmed quasars \citep{2020ApJS_SDSS_quasar,veron2010catalogue}, while outlier 324\footnote{DESI 39627781683806700} has been identified as a BL Lac object by \cite{ramazani2017empirical}.
Calibration issues, with either a single arm missing or different arms being improperly aligned (e.g. for outliers 4\footnote{DESI 39633149726034475}, 10\footnote{DESI 39633154058748211}, and 31\footnote{DESI 39633154041974595}), can also cause low probability estimates from both flow models.

\section{Discussion and Conclusions}
\label{sec:conclusion}

In this work, we have identified and cataloged outliers from the DESI BGS with a spectrum autoencoder and a normalizing flow. The overall high quality of DESI spectra allows us to present a rich collection of objects exhibiting unusual physical properties, ranging from irregular or double-peaked narrow emission lines, blended objects, galaxy mergers, to potential dual AGN in close separation, BALs, and ERQs. 
However, our outlier detection scheme has also unveiled issues related to the data collection and processing pipelines. 
Predominantly, we find BGS outliers to be stars, misclassified by {\em Redrock} as galaxies.

This finding prompted the development of our own star model, following the same approach as for galaxies, but now applied to stellar spectra from the DESI MWS.
We find good modeling quality without the PCA residuals in the current {\em Redrock} PCA models.
We also identified several more outliers that conform neither with our galaxy nor our star model.
These are often blended sources, quasars, or spectra with calibration issues.

To facilitate further studies, we release the full catalog of galaxy and star probabilities and reconstruction qualities for all BGS objects at our code repository\footnote{\url{https://github.com/pmelchior/spender}}. The pre-trained autoencoder and normalizing flow models are also accessible with instructions at that URL.

As possible extensions of the work presented here, the pre-trained StarAE and StarNF models could be employed to identify unusual stars in the MWS sample for follow-up investigations.
It is also entirely possible to create a star-galaxy classifier by running a spectrum through both of our pipelines and then compare the probabilities from each of the NFs to make classification decisions based on the relative likelihoods.

\section*{Acknowledgments}
\begin{acknowledgments}
We are grateful for the help of Michael Strauss in interpreting the weirdest spectra we discuss in this work.

This work was supported by the AI Accelerator program of the Schmidt Futures Foundation.
The authors are pleased to acknowledge that the work reported on in this paper was substantially performed using the Princeton Research Computing resources at Princeton University, which is a consortium of groups led by the Princeton Institute for Computational Science and Engineering (PICSciE) and Office of Information Technology's Research Computing.
\end{acknowledgments}

%

\facilities{KPO:4m, APO:2.5m, Gaia, WISE}


\software{
    \href{https://github.com/pmelchior/spender}{\texttt{spender}} \citep{melchior2022},
    \href{https://pytorch.org/}{\texttt{Pytorch}} \citep{paszke2019pytorch},
    \href{https://corner.readthedocs.io/}{\texttt{corner}} \citep{corner},
    \href{https://umap-learn.readthedocs.io/}{\texttt{umap}} \citep{sainburg2021parametric},
    \href{https://github.com/huggingface/accelerate}{\texttt{accelerate}} \citep{accelerate-Gugger2022},
    \href{https://github.com/aliutkus/torchinterp1d}{\texttt{torchinterp1d}},
    \href{https://github.com/bayesiains/nflows}{\texttt{nflows}} \citep{nflows-Durkan2020},
    \href{http://www.numpy.org}{\texttt{NumPy}} \citep{harris2020array},
    \href{https://www.astropy.org/}{\texttt{Astropy}} \citep{Astropy2022ApJ},
    \href{https://matplotlib.org}{\texttt{Matplotlib}} \citep{Hunter:2007}
}



\newpage
\bibliography{main}{}
\bibliographystyle{aasjournal}

\end{document}

%% file: data.tex
We use data from the DESI EDR, which includes spectra observed during the Survey Validation (SV) campaign that was conducted before the start of the main survey to evaluate DESI's scientific program~\citep{desi_sv}.
SV was divided into two main phases: an initial `Target Selection Validation' phase to finalize the target selection and a pilot survey of the full DESI program that covered $\sim$ 140 ${\rm deg}^2$, the One-Percent Survey. 

In this work we focus on targets from the BGS and the MWS. 
The initial target lists are defined on the basis of the imaging data from the Legacy Surveys~\citep[LS;][]{desi_ls}.
An object is considered a BGS target if it is either not in the {\em Gaia} Data Release 2 catalog \citep{gaia_dr2} or if it is in {\em Gaia} and has $(G_{Gaia} - r_{\rm raw}) > 0.6$,
where $G_{Gaia}$ is the $G$-band magnitude from {\em Gaia} and $r_{\rm raw}$ is the LS $r$-band magnitude without galactic extinction correction. 
Afterwards, a fiber-magnitude cut is imposed to remove imaging artifacts or other spurious objects. 
Quality cuts then remove any object without photometric observations in all three LS optical bands and with colors outside $-1 < (g-r) < 4$ and $-1  < (r -z) < 4$. 
Very bright objects with $r > 12$ and $r_{\rm fibertot} < 15$ are also removed from the sample.
For further details on the BGS target selection, see \cite{bgs_ts}. 

MWS targets consists of objects that are in both LS and {\em Gaia} Early Data Release 3 \citep{Gaia_EDR3}. 
They are restricted to objects that are classified as point sources based on their morphology and within the magnitude limits $16 < r < 19.2$ and $r_{\rm raw} < 20$. 
There is an additional cut on {\em Gaia} astrometric excess noise ({\em Gaia} AEN $< 3$) as well as quality cuts on the the LS $g$ and $r$-band photometry. 
No quality cuts on $z$-band photometry are imposed.
For further details on the MWS target selection, see \citet{mws_ts}. 

All spectra in the EDR are reduced using the `fuji' version of the DESI spectroscopic data reduction pipeline \citep{desi_spec_pipe}. 
First, spectra are extracted from the spectrograph CCDs using the 
{\em spectroperfectionism} algorithm of \cite{bolton2010}.
Then, fiber-to-fiber variations are corrected by flat-fielding, and a sky model,
empirically derived from sky fibers, is subtracted from each spectrum.
Afterwards, the fluxes in the spectra are calibrated using stellar model fits
to standard stars. 
The calibrated spectra are co-added across exposures to produce the final processed spectra. 
Among the different co-adds in the EDR, we use the ones produced by combining all available spectra for a given target. 

For each spectrum, the DESI EDR provides redshift measurements from the
{\em Redrock}\footnote{\url{https://redrock.readthedocs.io}} 
redshift fitting algorithm. 
For a given spectrum, {\em Redrock} derives redshift by minimizing the  $\chi^2$ between the observed spectrum and a model spectrum constructed from a linear combination of Principal Component Analysis (PCA) basis spectral templates in three template classes (``stellar'',  ``galaxy'', and ``quasar'').
The redshift and template class that produces the lowest $\chi^2$ is used as the best-fit redshift and spectral classification.
{\em Redrock} also provides an estimate of redshift uncertainties, $\mathtt{ZERR}$, and a redshift confidence measurements, $\Delta\chi^2$, which corresponds to the difference between the $\chi^2$ values of the best-fit model and the next best-fit model.

For BGS, we only use spectra observed with functioning fiber positioners and with reliable redshift measurements as defined in \cite{bgs_ts}. 
We only keep spectra classified as galaxy spectra by 
{\em Redrock} (i.e. {\tt SPECTYPE=="GALAXY"}) with redshifts $z\in[0,0.8]$ (cutting off a very minor high-redshift tail) and having no {\em Redrock} warning flags {\tt ZWARN}.
From MWS, we select spectra with {\tt ZWARN==0} and {\tt SPECTYPE=="STAR"}.
